\documentclass[a4paper]{article}
\usepackage[right=2.5cm,left=2.5cm,top=3.5cm,bottom=3.5cm,headsep=0cm,footskip=0.5cm]{geometry}
\usepackage{graphicx}
\usepackage{amssymb}
\usepackage{amsmath}
\usepackage{slashed}
\usepackage{color}

\title{\textbf{Cosmology with Scalar-Euler form Coupling}}
\author{Adolfo Toloza$^{1,2}$ and Jorge Zanelli$^{1,3}$\\
$^1$  {\small \textit{Centro de Estudios Cient\'{\i}ficos, Arturo Prat 514, Valdivia, Chile.}} \\
$^2$ {\small \textit{Instituto de F\'{\i}sica, P. Universidad Cat\'olica de Valpara\'{\i}so,  Casilla 4059, Valpara\'{\i}so, Chile.}} \\ 
$^3$ {\small \textit{Universidad Andr\'es Bello,  Av. Rep\'ublica 440, Santiago, Chile.}}}
\date{}

\begin{document}
\maketitle

\begin{abstract}
A coupling between the spacetime geometry and a scalar field involving the Euler four-form can have important consequences in General Relativity. The coupling is a four-dimensional version of the Jackiw-Teitelboim action, in which a scalar couples to the Euler two-form in two dimensions. In this case the first order formalism, in which the vierbein (or the metric) and the spin connection (or the affine connection) are varied independently, is not equivalent to the second order one, where the geometry is completely determined by the metric. This is because the torsion postulate $T^a\equiv 0$ is not valid now and one cannot algebraically solve the spin connection from its own field equation. The direct consequence of this obstruction is that the torsion becomes a new source for the metric curvature, and even if the scalar field is very slowly varying over cosmic scales so as to have no observable astronomical effects at the galactic scale, it has important dynamical effects that can give rise to a cosmological evolution  radically different from the standard FRWL model.
\end{abstract}

%%%%%%%%%%%%%%%%%%%%%
\section{Introduction}  % 1 %
%%%%%%%%%%%%%%%%%%%%%
A century after Einstein's formulation of General Relativity (GR) no one doubts that gravitation is a manifestation of the geometry of spacetime: matter and energy curves spacetime on which matter and energy move. It is by looking at the trajectories of particles and light that one can determine the spacetime geometry and thereby infer the local energy-matter distribution. Similarly, by studying the shape and evolution of the universe one infers the distribution of energy and matter at the cosmic scale. Thus, at the galactic scale, the velocity distribution of stars suggests the presence of (dark) matter of an unknown nature; at the cosmic scale, the accelerated expansion of the universe points towards the existence of a form of vacuum (dark) energy with exotic properties ($p<-1/3\rho$). 

These puzzling results have piled up over the past decades on top of long-standing riddles like the unexpected smallness of the cosmological constant and the quantum understanding of gravity. Even if these conclusions were not the result of inaccurate measurements, or wrong assessments of the data, it might be a healthy attitude to consider variations on the assumptions of the theoretical framework and to explore the scenarios resulting from such variations.

%%%%%%%%%%%%%%%%%%%%%%%%%%%%%%%%%%%%%%%%%
\subsection{The standard cosmological model}  % 1.1 %
%%%%%%%%%%%%%%%%%%%%%%%%%%%%%%%%%%%%%%%%%

The current view is that the universe is roughly described by the standard Friedmann-Robertson-Walker-Lemaitre model, also referred to as big-bang cosmology. The main ingredients in this model are: \\
\
$\bullet$ Symmetry (principle of equivalence): Spacetime is a differentiable manifold endowed with a bundle of tangent spaces and the laws of physics are invariant under local Lorentz transformations \cite{O'R-S,Z,zReview}.\\
$\bullet$ Dynamics: Extremizing the Einstein-Hilbert action
\begin{equation}
I[g] = \int_M \sqrt{|g|}\Big[\frac{1}{8\pi G}(R -2\Lambda )\Big]\mathrm{d}^4x + \int_M L_{matter}\mathrm{d}^4x, \label{EH}
\end{equation}
with respect to the metric $g_{\mu \nu}$, yields Einstein's equations (\textbf{EE}), 
\begin{equation}\label{EE}
R^{\mu}_{\, \, \nu} -\frac{1}{2} R\delta^{\mu}_{\, \, \nu}  + \Lambda \delta^{\mu}_{\, \, \nu} =8\pi GT^{\mu}_{\, \, \nu}\, .
\end{equation} \\
 $\bullet$ Cosmological principle: The spacetime metric is a solution of EE that admits a global slicing into homogeneous and isotropic spacelike three-surfaces.\\
 
The validity of this model is supported on astronomical observations which seem consistent with the following facts: $13.7 \times 10^9$ years ago, the universe was hot, small, fairly homogeneous and isotropic; the spatial section of the universe is currently quite flat and homogeneous at large scales; the energy density of the universe is very near the critical value to make its spatial sections flat ($k=0$); the expansion of the universe seems to be accelerating. 

The theoretical challenge is to explain these observations with a consistent dynamical model. 

In order to model dark energy and/or dark matter the simplest option may be to include some exotic fluids that contribute to the right hand side (RHS) of the EE  (see, e.g., \cite{Gibbons:2003yj}). A second alternative could be to change the dynamics of geometry, defined by left hand side (LHS) of EE, replacing the Einstein-Hilbert Lagrangian by some scalar density function of the metric, the curvature and  the torsion $f(g,R,T)$.\footnote{For arbitrary $f$, this in general brings in new degrees of freedom for the graviton through higher order derivatives (see for instance \cite{de-Felice,Capozziello,Ferraro}), but an equally serious concern is the fact that there seems to be no principle from which the function $f$ can be derived. This problem is like that of the cosmological constant but not just for one arbitrary parameter but for infinitely many.} A third option would be to relax the assumption of homogeneity and isotropy of the spatial sections (cosmological principle). In the absence of those assumptions, a cosmological model makes sense if it describes the evolution of averaged geometrical and physical properties and this may considerably change the conclusions see e.g., \cite{Andersson:2011za}. Here, however, we will not consider this option, and stick to the conventional simplifying assumptions of homogeneity and isotropy.

In most of the explorations described above a clear distinction between what goes into LHS (geometry) and RHS (matter-energy content) of (\ref{EE}) is implicitly assumed. This scheme is a direct extension of the experience in classical mechanics and electrodynamics, where the sources (RHS) determine the evolution of the relevant dynamical variables --positions and field configurations-- (LHS). This point of view is so ingrained in our way of thinking about spacetime that we infer the matter content of the universe from the motion of particles. If the geodesics are straight lines, we assume spacetime to be flat and there is no matter anywhere. Conversely, if the geodesics are not straight lines, we infer that there must be matter somewhere, even if invisible. This, however, is not necessarily true. It has been shown, for instance, that there exist forms of matter coupled to gravity that do not necessarily curve spacetime \cite{AMTZ}.  

Here we show that the converse may also fail to be true: there could be sources for local curvature that are not necessarily forms of matter; the source of curvature might be geometry itself, and this may give rise to a radically different cosmological evolution. As we shall see next, a situation like this may result from a minimal modification of GR, showing how sensitive on the assumptions of GR the resulting cosmology can be.
  
%%%%%%%%%%%%%%%%%%%%%%%%%%%%%%%%%%%%%%%%%%%
\subsection{Second order formalism and torsion}   % 1. 2 %
%%%%%%%%%%%%%%%%%%%%%%%%%%%%%%%%%%%%%%%%%%%

Another important simplifying assumption postulated by Einstein in the standard approach to GR is the vanishing of torsion. According to this condition, the affine connection is symmetric and defined by the Christoffel symbols, so the metric is the only fundamental field to be varied in the action. The Einstein-Hilbert action is a functional of the metric and varying it yields second order equations for $g_{\mu \nu}$.

On the other hand, in the first order formalism the action is varied with respect to the vielbein $e^a$ and the Lorentz connection $\omega^{ab}$independently, in a similar scheme to that proposed by Palatini \cite{Palatini}. In this case, since $e^a$ and  $\omega^{ab}$ are assumed to be independent, the torsion two-form $T^a=de^a+\omega^a{}_b e^b$ cannot identically vanish, but it must be determined by the field equations. In four-dimensional matter-free gravity one of the equations in the first order approach implies $T^a=0$. Hence, the two formalisms are equivalent in this case, because the torsion-free condition is always satisfied: either as an on shell condition (first order approach), or as an identity (second order formalism).

Moreover, in the four-dimensional, matter-free case, the equivalence also holds off-shell, because the equation $T^a=0$, obtained by varying the action with respect to $\omega^{ab}$, can be \textit{algebraically} solved for $\omega$ as a function of $e^a$. By the implicit function(al) theorem, this means that the reduced action obtained by substituting $\omega=\omega(e)$, the action $\tilde{I}[e]=I[e, \omega(e)]$ yields a second order action that is completely equivalent to the original first order one, even away from the classical extrema \cite{Henneaux-Teitelboim}. Thus, in four-dimensional matter-free GR, the torsion-free condition does not restrict the dynamics around classical solutions (see, e.g., \cite{Contreras:1999je,Aros:2003bi}). On the other hand, if the torsion equation, obtained varying with respect to $\omega$, cannot be algebraically solved for $\omega$ as a function of the other fields, the first and second order formalisms are not equivalent.

If $T^a\neq 0$, the presence of torsion could be revealed by fermions, as they would follow different geodesics from the those for spin-0 or spin-1 particles, an  effect that may not be significant for the current experimental accuracy (see, e.g., \cite{Carroll-Field}).  However, even though postulating $T^a\equiv 0$ might seem harmless and could pass the experimental tests in four dimensions, it is not a logical necessity and it is theoretically unsatisfactory to impose $T^a= 0$, a constraint that is not respected in the presence of ordinary matter, like spin 1/2 fields. 

Another, perhaps more important aspect of torsion, is that it may act as a source for the metric curvature. This may have important cosmological consequences. The purpose of this paper is to present an example of this effect.

%%%%%%%%%%%%%%%%%%%%%%%%%%%%%%%%%%%%%%%
\subsection{The scalar-Euler form coupling}  % 1. 3 %
%%%%%%%%%%%%%%%%%%%%%%%%%%%%%%%%%%%%%%%

The modification we consider here consists of adding a coupling term between a scalar field and the Euler density,
\begin{equation}\label{coupling}
\phi \epsilon_{abcd}R^{ab}\wedge R^{cd},
\end{equation} 
where $R^{ab} =\mathrm{d}\omega^{ab} + \omega^a{}_c \wedge \omega^{cb}$ is the curvature two-form for the Lorentz connection $\omega^a{}_b$.  For constant $\phi$ the modification (\ref{coupling}) is irrelevant, as it adds a constant to the action proportional to the Euler characteristic of spacetime. So, one can expect that if $\phi$ is slowly varying over cosmological distances, the Einstein equations should not be affected at planetary scales. However, even if at present the effects of the modification could be negligible, the additional term changes the field equations by adding a new equation, $\epsilon_{abcd}R^{ab}\wedge R^{cd}=0$, and modifying the other equations in a form that could significantly affect the evolution near the big-bang as well as at the final stages of the universe's expansion. 

The scalar $\phi$ could be an effective field whose origin is not relevant for the present discussion. With the possible exception of the Brout-Englert-Higgs particle, all scalars in nature are composites of spin-1/2 matter fields, whose origin could be found in the underlying fundamental microscopic theory. Still other possibilities exist which will be discussed in Section 4.

The scalar coupling to the Euler form can also be found in the dimensional compactification of a higher-dimensional gravitational theory of the Lovelock family, where the first nontrivial correction to the Einstein-Hilbert action in dimensions $\mathrm{D} \geq 5$ is the Gauss-Bonnet term. For instance, in five dimensions this term is
\begin{equation}
 \epsilon_{ABCDF}R^{AB}\wedge R^{CD}\wedge e^F.
\end{equation}  
Reducing to four dimensions, and using 
\begin{equation}
e^A_{\mu}=\left(
\begin{array}{ll}
 e^a_{\mu} & A_\mu \\
 \varphi^a & \phi
\end{array}
 \right),
\end{equation}
produces (\ref{coupling}) \cite{VanAcoleyen} along with other non-minimal couplings involving torsion explicitly \cite{Mardones}.

The scalar-Euler density coupling (\ref{coupling}) is the natural extension to four dimensions of the two-dimensional Jackiw-Teitelboim gravitation theory \cite{Teitelboim-2DGravity,Jackiw-2DGravity}, whose gravitational Lagrangian is the product of a scalar field and the two-dimensional Euler form, $\phi\epsilon_{ab}R^{ab}$.

This coupling could also be seen as a particular choice of the most general extension of scalar-tensor theories originally proposed by Horndeski \cite{Horndeski}, recently revived in the context of Galileons \cite{Deffayet:2009de,Deffayet:2011gz,Charmousis}, where the combination $V(\phi)(R^2-4R^{\mu \nu}R_{\mu \nu} + R^{\alpha \beta \mu \nu} R_{\alpha \beta \mu \nu})$ is one of the possible extensions of gravity that yields second order field equations for the metric and $\phi$. 

The main difference between our work, originally discussed in \cite{Rio2011}, and the Horndeski approach is that we adopt the first order formalism where $e^a$ and $\omega^{ab}$ are dynamically independent fields. This makes a great difference because in the presence of scalar-Euler form coupling the first and second order formalisms are no longer equivalent. The contribution of (\ref{coupling}) to the variation of the action with respect to $\omega^{ab}$ implies that torsion does not vanish if $\phi$ is not constant. Moreover, this equation is not algebraically solvable for the connection as a function of $e^a$ and $\phi$. Hence, there is no local expression $\omega^{ab}(e,\phi)$ to substitute in the action. Thus, the presence of the term (\ref{coupling}) prevents writing the action only in terms of the vielbein (or the metric) and $\phi$. 

On the other hand, in the second order formalism, the field equation obtained by varying $\phi$, means the vanishing of the Gauss-Bonnet invariant, $\epsilon_{abcd}R^{ab}\wedge R^{cd}=0$ which, for a 4D Einstein space, $R_{\mu \nu}=s(x)g_{\mu \nu}$, implies the vanishing of the  Kretschmann invariant, $R^{\alpha \beta \mu \nu}R_{\alpha \beta \mu \nu}=0$. This equation represents a strong constraint on the geometry, incompatible with most classical solutions of GR, such as the Kerr black hole, which rules out a coupling like (\ref{coupling}) in the second order formalism, unless additional modifications to the theory are introduced. This is not the case in the first order formalism since the condition (\ref{Euler}) below, establishes a relation between purely metric part of the curvature and torsion which, as we shall see in section 3, it does not necessarily eliminate interesting options.

%%%%%%%%%%%%%%%%%%%%%%%%%%%%%%%%%%%%%%%
\section{First order action and field equations}  % 2 %
%%%%%%%%%%%%%%%%%%%%%%%%%%%%%%%%%%%%%%%

Let us consider the first order form of the Einstein-Hilbert action with cosmological constant in four-dimensions, including the scalar-Euler density coupling (\ref{coupling}).\footnote{Henceforth wedge products of forms will be assumed.} We will also allow the possible presence of other forms of non geometric matter represented by some unspecified fluid, whose dynamics is not relevant for the present discussion. The action reads
\begin{equation}\label{fullaction}
I[e,\omega,\phi] = \frac{1}{32 \pi G}\int \epsilon_{abcd} \Big(R^{ab} e^c e^d-\frac{\Lambda}{6} e^a e^b e^c e^d + \phi R^{ab} R^{cd} \Big) + \int L_m
\end{equation}
where $\Lambda$ is the cosmological constant, and $L_m$ represents the non geometric matter. Varying with respect to $e^a$, $\omega^{ab}$ and $\phi$, yields the following field equations
\begin{eqnarray}
\label{Einstein} \delta e &:& \epsilon_{abcd} \Big(R^{ab} - \frac{\Lambda}{3} e^a e^b \Big) e^c = 16 \pi G \tau_d \\
\label{Torsion} \delta \omega &:& \epsilon_{abcd} (T^a e^b + \mathrm{d}\phi R^{ab}) = 0 \\
\label{Euler} \delta \phi &:& \epsilon_{abcd} R^{ab} R^{cd} = 0 \, .
\end{eqnarray}
where $\tau_d$ is the stress-energy 3-form\footnote{The stress-energy 3-form is related to the stress-energy tensor as $*\tau_d = \tau_{cd} e^c$, where ${\tau^c}_d = |e|^{-1} e^c_\mu \frac{\delta(|e| L_m)}{\delta e^d_\mu}$}
 defined by $\delta L_m = \delta e^d \wedge \tau_d$. Eq. (\ref{Einstein}) has the same form as the Einstein equations with the important difference that since now (\ref{Torsion}) implies that torsion can no longer be assumed to vanish, $R^{ab}$ is not the usual Riemann curvature of standard GR. This is best seen splitting the Lorentz connection as $\omega^{ab}=\bar{\omega}^{ab}+\kappa^{ab}$, where $\bar{\omega}^{ab}$ is the torsion-free part,
\begin{equation}
\mathrm{d} e^a + \bar{\omega}^a_{\,\,\,\,b} \wedge e^b =0 \rightarrow \bar{\omega}^a_{\,\,\,\,b \mu}=e^a_\nu (\nabla_{\mu} E_b^\nu), \label{0-torsion}
\end{equation}
and the contorsion $\kappa^{ab}$ determines the torsion, $T^a = {\kappa^a}_{b} \wedge e^b$. In (\ref{0-torsion}), $E_a^\nu$ is the inverse vielbein and $\nabla$ is the covariant derivative in the Christoffel connection. In this scheme, the Lorentz curvature splits into de usual torsion-free part of Riemannian geometry, $\bar{R}^{ab} $, and the torsion-dependent terms,
\begin{equation}
\label{Curvature} R^{ab} = \bar{R}^{ab} + \bar{\mathrm{D}} \kappa^{ab} + {\kappa^a}_c \wedge \kappa^{cb},
\end{equation}
where $\bar{\mathrm{D}}$ is the covariant derivative in the torsionless conection $\bar{\omega}^{ab}$. In the presence of torsion the $\kappa$-terms become sources for the Riemann curvature $\bar{R}^{ab}$ in the Einstein equations.  Using (\ref{Curvature}), Eq.(\ref{Einstein}) becomes
\begin{equation}
\epsilon_{abcd} \Big(\bar{R}^{bc} -\frac{\Lambda}{3} e^b e^c \Big) e^d = -\epsilon_{abcd} (\bar{\mathrm{D}} \kappa^{bc} + {\kappa^b}_f \kappa^{fc}) e^d + 16\pi G \tau_a\, , \label{newEE}
\end{equation}
where the torsion terms on the RHS add to the other matter sources represented by $\tau_a$. 

In the metric formulation, the field equations for the same action would have been completely different. The metric expression corresponding to (\ref{coupling}) is 
\begin{equation}
\phi\Big(R^2- 4 R^{\mu \nu} R_{\mu \nu}  + R^{\mu \nu \alpha \beta} R_{\mu \nu \alpha \beta} \Big)\sqrt{-g},
\end{equation}
which has been noted to give rise to second order field equations  \cite{Horndeski,Charmousis}. However, varying with respect to $g_{\mu \nu}$ under the assumption of vanishing torsion yields, instead of  (\ref{Einstein}) and (\ref{Torsion}), the Einstein equations plus quadratic curvature corrections and terms containing both first and second order derivatives of $\phi$. It can be easily seen that the resulting sets of equations of both formulations are inequivalent.

%%%%%%%%%%%%%%%%%%%%%%%%%%%%%%%%%%
\section{Cosmological consequences}    % 3  %
%%%%%%%%%%%%%%%%%%%%%%%%%%%%%%%%%%

We now consider a spacetime foliated by a family of isotropic and homogeneous three-dimensional spatial slices (Bianchi types I, V \& IX), as described by the standard Big-Bang cosmology, and we examine the solutions of the field equations (\ref{Einstein} -- \ref{Euler}) in this case. The metric, 
\begin{equation} \label{g-mu-nu}
\mathrm{d}s^2 = -\mathrm{d}t^2 + a^2(t)\Big[\frac{\mathrm{d}r^2}{1-kr^2}+r^2(\mathrm{d}\theta^2 + \sin^2\theta \mathrm{d}\varphi ^2)\Big],
\end{equation}
admits six global Killing vectors ($\pounds_\xi g_{\mu \nu} = 0$), associated to three spatial translations, $\xi_{(i)}$, and three rotations, $\xi_{(ij)}$, where\footnote{The usual relation between cartesian $(x_1,x_2,x_3)$ and polar $(r,\theta,\varphi)$ coordinates is understood.}
\begin{equation} \label{Killing}
\xi_{(i)} = \sqrt{1-kr^2} \partial_{x_i}\; , \;\;  \mbox{and}\;\; \; \xi_{(ij)}=x_i \partial_{x_j} - x_j \partial_{x_i}.
\end{equation}
We assume torsion and the scalar field to have the same isometries as the background spacetime, $\pounds_\xi {T^{\alpha}}_{\mu \nu} = 0$ and  $\pounds_\xi \phi = 0$. It is straightforward to see that these conditions imply $\phi=\phi(t)$, and the only nonvanishing components of ${T^{\alpha}}_{\mu \nu}$ are 
\begin{eqnarray} 
\nonumber {T^r}_{\theta \varphi} = 2 f(t) a(t) r^2 \sqrt{1-kr^2} \sin\theta &;& {T^\varphi}_{r \theta} = \frac{2 f(t) a(t)}{\sqrt{1-kr^2}\sin \theta} \\
\label{Ta-mu-nu} {T^\theta}_{r \varphi} = - \frac{2 f(t) a(t) \sin \theta}{\sqrt{1-kr^2}} &;& {T^r}_{r t} = {T^\theta}_{\theta t} = {T^\varphi}_{\varphi t} = h(t)\; ,
\end{eqnarray}
where $h(t)$ and $f(t)$ are functions of time to be determined by the field equations. 

As shown in Appendix A, substituting the expressions for the metric an torsion into the first order field equations (\ref{Einstein} -- \ref{Euler}), yields
\begin{eqnarray}
\label{1} &&U^2 + \frac{k}{a^2} - f^2 - \frac{\Lambda}{3} = \frac{\kappa^2}{3}\rho_m \\
\label{2} &&2(\dot{U} + H U) + U^2 + \frac{k}{a^2} - f^2 - \Lambda = -\kappa^2 p_m \\
\label{3} &&-h + \dot{\phi} \Big(U^2 + \frac{k}{a^2} - f^2\Big) = 0\\
\label{4} &&\Big(-\frac{1}{2} + \dot{\phi} U \Big) f = 0\\
\label{5} &&\Big(U^2 + \frac{k}{a^2} - f^2\Big)(\dot{U} + H U) - 2 f U(\dot{f} + H f) = 0
\end{eqnarray}
where $H=\dot{a}/a$ is the Hubble function, $\kappa^2 = 8 \pi G$, $\rho_m$ and $p_m$ are the energy density and pressure described by the matter Lagrangian, and $U \equiv H + h$.  Taking the covariant derivative of (\ref{Einstein}) the stress 3-form is found to satisfy the on-shell condition $\mathrm{D}\tau_d = {T^c}_{d l} e^l\tau_c$ or equivalently,
\begin{equation}\label{conservation}
\mathrm{\bar{D}}\tau_d = {\kappa^c}_{ld} e^l \tau_c \; ,
\end{equation}
were $\bar{\mathrm{D}} = \mathrm{d} + \bar{\omega}$ denotes the covariant derivative in the torsion-free connection.   As seen in the appendix, here $\kappa^{0I} = h(t) e^I$ and $\kappa^{IJ} = -f(t) \epsilon^{IJ}{}_K e^K$, which implies that the right hand side of (\ref{conservation}) vanishes identically. This implies the continuity equation for the matter content of the universe, $\dot{\rho}_m + 3 H (\rho_m + p_m) = 0$, which is ultimately related to the fact that the metric (\ref{g-mu-nu}) and the torsion (\ref{Ta-mu-nu}) satisfy Bianchi identities $\mathrm{D} R^a{}_b = \mathrm{d}R^a{}_b + \omega^a{}_c R^c{}_b - R^a{}_c \omega^c{}_b \equiv 0$, and $\mathrm{D} T^a =R^a{}_c e^c$.\footnote{This is not an obvious property of our ansatz, because the expression (\ref{Ta-mu-nu}) is only a consequence of the tensorial character of $T^\alpha{}_{\mu\nu}$ in a space of that admits the Killing vectors (\ref{Killing}) and does not depend on $T^\alpha{}_{\mu\nu}$ being the torsion tensor.}

 Equations (\ref{1}) and (\ref{2}) correspond to the modified Friedmann equations and can be cast in the standard form: $H^2 +\frac{k}{a^2} - \frac{\Lambda}{3} = \frac{\kappa^2}{3} \rho$ and $2\dot{H} + 3H^2 +\frac{k}{a^2} - \Lambda = -\kappa^2 p$, where $\rho= \rho_m  + \rho_T$ and $p=p_m + p_T$ combine the density and pressure of ordinary matter and the effect of torsion. Thus, the effective density and pressure produced by torsion can be identified as 
\begin{eqnarray} \label{rho}
\rho_T &=&-\frac{3}{\kappa^2}(2 H h + h^2 - f^2) , \\
\label{p}
p_T &=& \frac{1}{\kappa^2}(2\dot{h} + 4 H h + h^2 - f^2).
\end{eqnarray} 
This identification is made more compelling by the fact that $\rho_T$ and $p_T$ can be shown to satisfy the corresponding continuity equation, $\dot{\rho}_T + 3 H (\rho_T + p_T) = 0$.\\

%%%%%%%%%%%%%%%%%%%%%%%%%%%%%%%%%%%%%%%%%%%%
\subsection{Solutions for pressureless matter}  % 3.1 %
%%%%%%%%%%%%%%%%%%%%%%%%%%%%%%%%%%%%%%%%%%%%

We consider ordinary matter described by dust $p_m=0$. The continuity equation $\dot{\rho}_m + 3 H \rho_m = 0$, implies $\rho_m = \rho_0 (a_0/a)^3$, where $\rho_0$ represents the present density of the universe, and $a_0$ is the present value of the scale factor which can be normalized to one. In order to solve the system, we first note that since equation (\ref{5}) represents the Euler invariant \cite{Nakahara}, it is locally a total derivative. Indeed, (\ref{5}) can be written as 
\begin{equation}
\frac{d}{dt}\Big(a^3\Big[\frac{U^3}{3} + U \Big(\frac{k}{a^2} - f^2\Big)\Big]\Big)=0 \; ,
\end{equation}
whose integral is
\begin{equation}
\label{6} \frac{U^3}{3} + U \Big(\frac{k}{a^2} - f^2\Big) = \frac{C}{a^3} ,
\end{equation}
where $C$ is an integration constant. Assuming $f\neq 0$ in (\ref{4}) implies $\dot{\phi}=(2U)^{-1}$, and replacing (\ref{3}) in (\ref{2}) allows integrating $U(t)$ as
\begin{equation}
U(t) = \sqrt{\frac{\Lambda}{2}} \tanh\Big[\sqrt{\Lambda/2} (t-t_0)\Big],
\end{equation}
for $\Lambda > 0$, where $t_0$ is an integration constant that sets the origin of time. For $\Lambda < 0$ the solution is
\begin{equation}
U(t) = -\sqrt{\frac{-\Lambda}{2}} \tan\Big[\sqrt{-\Lambda/2} (t-t_0)\Big],
\end{equation}
and for $\Lambda = 0$
\begin{equation}
U(t) = (t-t_0)^{-1}.
\end{equation}
Note that these results do not depend on $\rho_0$ or $k$. Hence, from (\ref{1}), (\ref{4}) and (\ref{6}) $a$, $\dot{\phi}$ and $f$ are obtained, and $h$ can be finally  read from (\ref{3}).  The solutions are different for each sign of $\Lambda$: \\

\textbf{i)$\Lambda > 0$}\\
\begin{eqnarray}
a(t) &=& \frac{1}{\sqrt{\Lambda}} \cosh\tau \Bigg(\frac{3 \sqrt{2} C}{\sinh\tau}-\frac{\sqrt{\Lambda} \kappa^2 \rho_0}{\cosh\tau}\Bigg)^{1/3} \\
\phi(t) &=& \frac{1}{\Lambda}\log\Big|\sinh\tau\Big| - \phi_0 \\
h(t) &=& \frac{1}{3} \sqrt{\frac{\Lambda}{2}} \coth\tau \Bigg(\frac{3 \sqrt{2} C - \sqrt{\Lambda} \kappa^2 \rho_0 \tanh^3\tau}{3\sqrt{2} C - \sqrt{\Lambda} \kappa^2 \rho_0 \tanh\tau}\Bigg) \\
f(t) &=& \sqrt{\Lambda}\Bigg(\frac{k \sinh^{2/3}\tau}{\cosh^{2}\tau \big(3 \sqrt{2} C - \sqrt{\Lambda} \kappa^2 \rho_0 \tanh\tau\big)^{2/3}} - \frac{2C-3C\tanh^2\tau+\frac{\sqrt{2\Lambda}}{6} \kappa^2 \rho_0 \tanh^{3}\tau}{6 C - \sqrt{2 \Lambda} \kappa^2 \rho_0 \tanh\tau}\Bigg)^{1/2}
\end{eqnarray}\\

\textbf{ii)$\Lambda < 0$}
\begin{eqnarray}
a(t) &=& \frac{1}{\sqrt{-\Lambda}} \sin\tau \Bigg(-\frac{3 \sqrt{2} C}{\cos\tau}+\frac{\sqrt{-\Lambda} \kappa^2 \rho_0}{\sin\tau}\Bigg)^{1/3} \\
\phi(t) &=& \frac{1}{\Lambda}\log\Big|\cos\tau\Big| + \phi_0 \\
h(t) &=& \frac{1}{3} \sqrt{\frac{-\Lambda}{2}} \tan\tau \Bigg(\frac{-3 \sqrt{2} C - \sqrt{-\Lambda} \kappa^2 \rho_0 \cot^3\tau}{-3 \sqrt{2} C + \sqrt{-\Lambda} \kappa^2 \rho_0 \cot\tau}\Bigg) \\
f(t) &=& \sqrt{-\Lambda}\Bigg(\frac{k \csc^{2}\tau \cos^{2/3}\tau}{\big(3 \sqrt{2} C - \sqrt{-\Lambda} \kappa^2 \rho_0 \cot\tau\big)^{2/3}} + \frac{2C-3C\cot\tau-\frac{\sqrt{-2\Lambda}}{6} \kappa^2 \rho_0 \cot^{3}\tau}{6 C - \sqrt{-2 \Lambda} \kappa^2 \rho_0 \cot\tau}\Bigg)^{1/2}
\end{eqnarray}\\

\textbf{iii)$\Lambda = 0$}
\begin{eqnarray}
a(t) &=& \frac{1}{2^{2/3}}t^{2/3}(\kappa^2 \rho_0 - 3C t)^{1/3}\\
\phi(t) &=& \frac{1}{4} t^2 + \phi_0 \\
h(t) &=& \frac{\kappa^2 \rho_0}{3t (\kappa^2 \rho_0 - 3 C t)}\\
f(t) &=& \Bigg(\frac{2^{2/3}k}{t^{4/3} (\kappa^2 \rho_0 - 3 C t)^{2/3}} + \frac{\kappa^2 \rho_0 - 9 C t}{3 t^2 (\kappa^2 \rho_0 - 3 C t)}\Bigg)^{1/2}
\end{eqnarray}
where $\tau = \sqrt{\frac{|\Lambda|}{2}} \ t$ is the rescaled dimensionless time. We have fixed the origin of the time coordinate $t_0$ in such a way that, for $\Lambda \leq 0$ $a(0) = 0$ and for $\Lambda > 0$ $a(\tau \rightarrow 0) \rightarrow \infty$.\\
\begin{figure}
\centering
\includegraphics[width=0.4\textwidth]{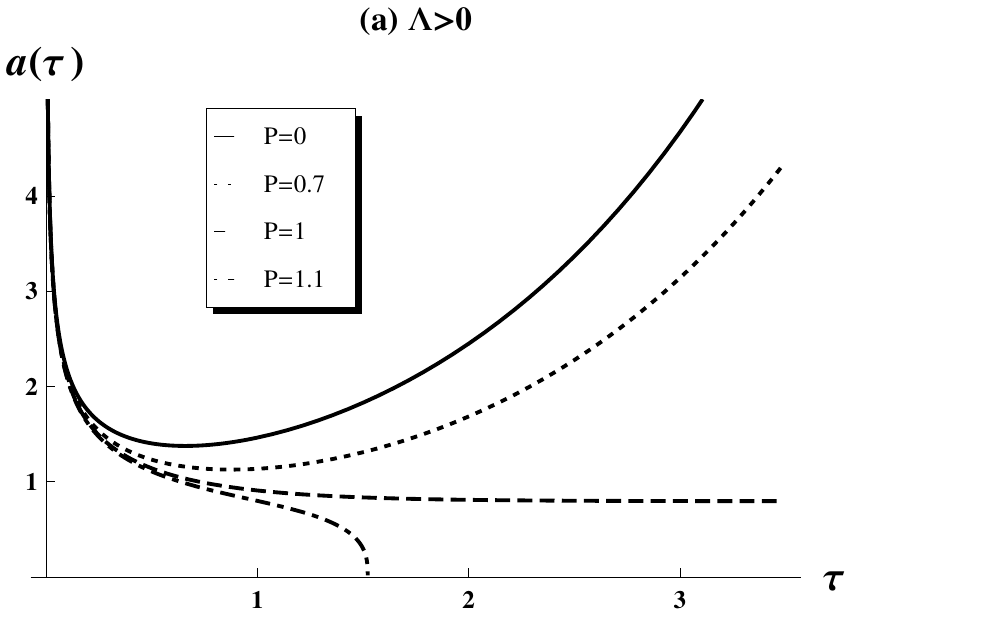}
\includegraphics[width=0.4\textwidth]{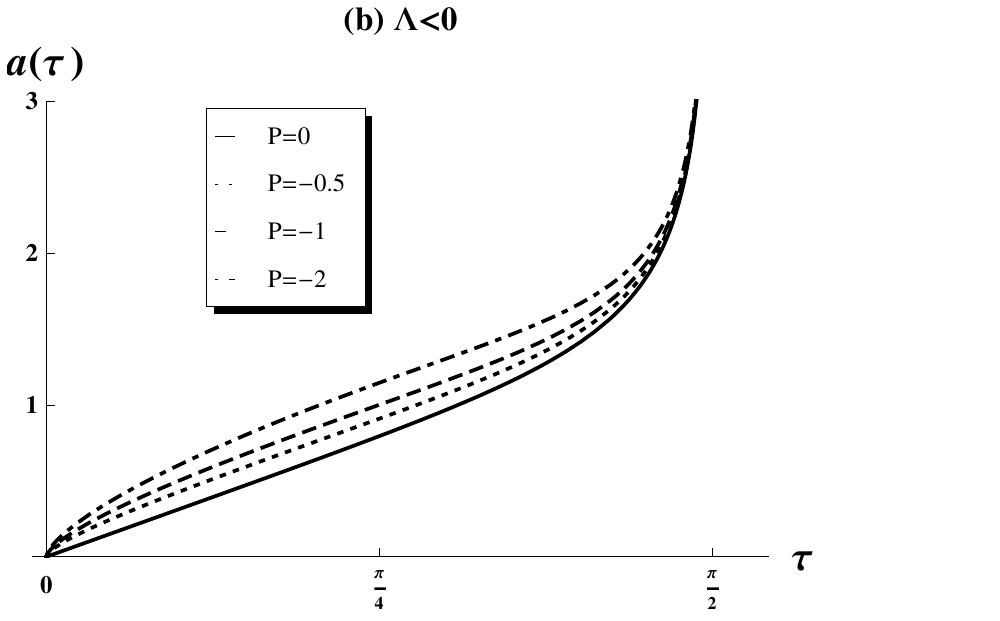}
\includegraphics[width=0.4\textwidth]{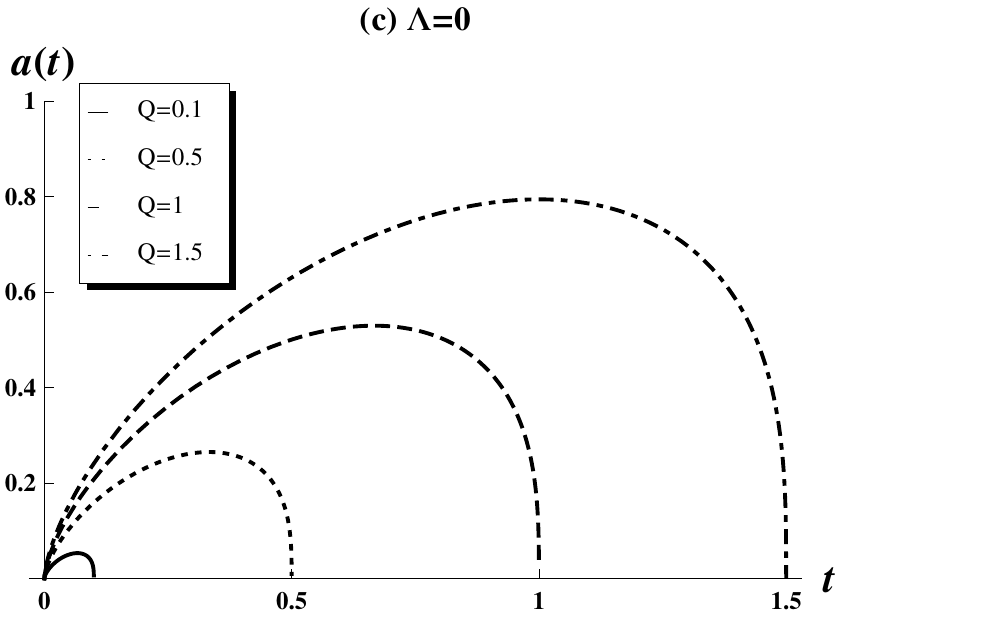}
\caption{Scale factor for $\Lambda$ positive, negative and zero respectively. For $\Lambda \neq 0$, $a(t)$ is in units of $|\Lambda|^{-1/2}$, and for $\Lambda = 0$ $a(t)$ has the same units as $t$. The graphs show the behaviour of $a(t)$ for different values of the dimensionless parameter $P=\sqrt{\frac{|\Lambda|}{2}}\frac{\kappa^2 \rho_0}{3 C}$ for the case $\Lambda \neq 0$, or for $\Lambda = 0$ the parameter $Q = \frac{\kappa^2 \rho_0}{3 C}$ has the same units as $t$}.
\label{scalefactor}
\end{figure}
\begin{figure}
\centering
\includegraphics[width=0.4\textwidth]{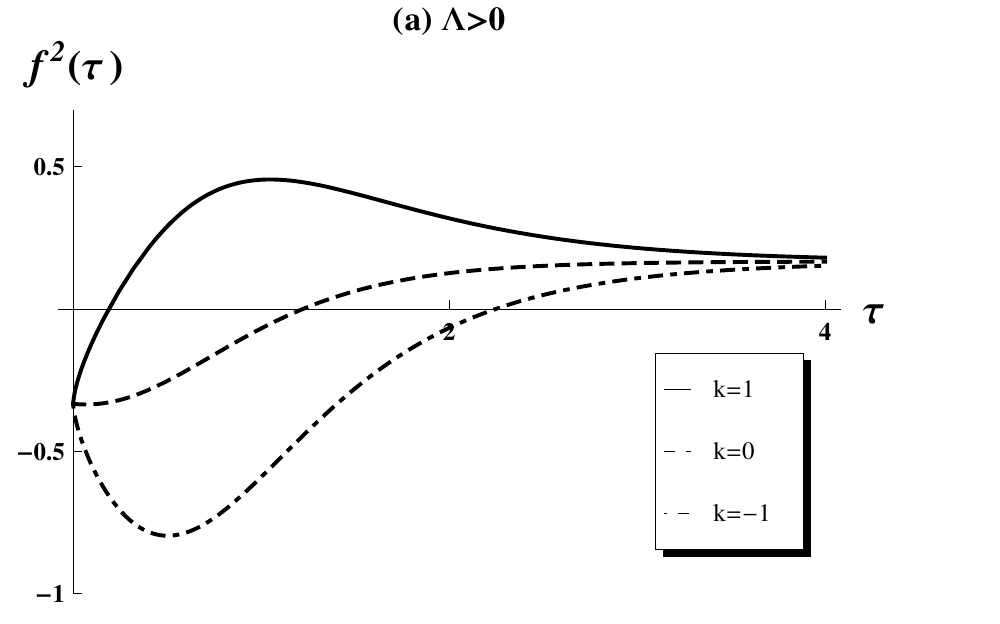}
\includegraphics[width=0.4\textwidth]{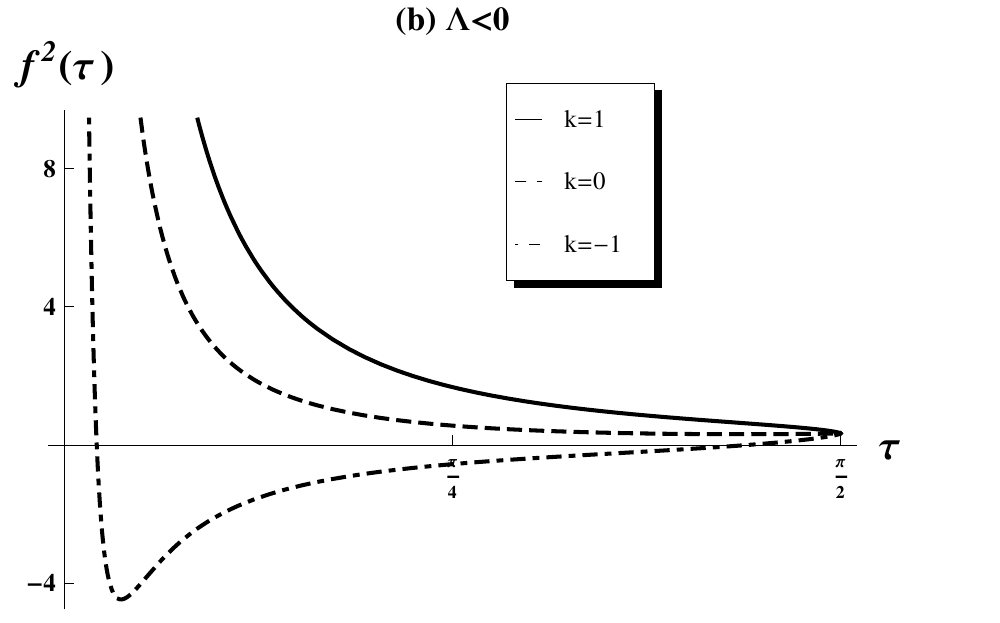}
\includegraphics[width=0.4\textwidth]{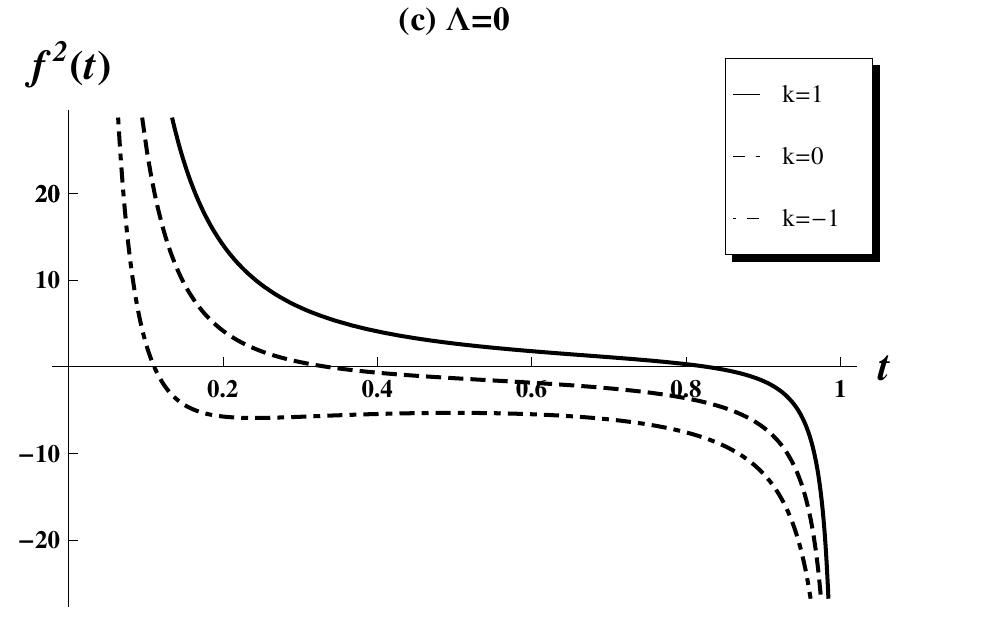}
\caption{Function $f^2(t)$ for $\Lambda$ positive, negative an zero respectively. For $\Lambda \neq 0$, $f^2(t)$ has units of $|\Lambda|$, for $\Lambda = 0$ $f^2(t)$ has units of $t^{-2}$. The graphs show the behaviour of $f^2(t)$ for different values of the spatial curvature $k$ and for particular values of the parameters $\rho_0$, $C$, and $\Lambda$. Continuous line corresponds to the case $k=1$, dashed lines is for $k=0$, and  dotdashed line for $k=-1$. Note that, in general, there are some values of $t$ for which $f^2(t)$ becomes negative.}
\label{fsquare}
\end{figure}

In Fig. (\ref{scalefactor}) we observe that for $C \neq 0$, for $\Lambda > 0$ the scale factor decreases from an infinite value, reaching a minimum and expanding again if the dimensionless parameter $P = \sqrt{\frac{\Lambda}{2}}\frac{\kappa^2 \rho_0}{3 C} < 1$, or collapsing at a finite time if $P > 1$; in the limiting case $P = 1$, the scale factor tends asymptotically to a finite constant. For $\Lambda \leq 0$ there is an initial singularity, if $\Lambda < 0$ there is an infinite expansion at a finite cosmic time, while with $\Lambda = 0$ there is always a Big-Crunch at $t_c = \frac{\kappa^2 \rho_0}{3 C}$; note that, counterintuitively, the time of collapse $t_c$  grows with the density. 

\newpage
The special case $C=0$ must be treated separately:\footnote{The sign of $a$ is irrelevant since only $a^2$ appears in the metric.}\\
a) For $\Lambda>0$:\\
$\bullet \;\;\; a \sim \cosh^{2/3}\tau$\\
$\bullet \;\;\;  h \sim \tanh\tau$\\
 $\bullet \;\;\;  f^2 \sim k\cosh^{-4/3}\tau + \tanh^2\tau$.\\
b) For $\Lambda<0$:\\
$\bullet \;\;\; a \sim \sin^{2/3}\tau$\\
$\bullet \;\;\;  h \sim \cot\tau$\\
 $\bullet \;\;\; f^2 \sim k\sin^{-4/3}\tau + \cot^2\tau$.\\
c) For $\Lambda=0$:\\
$\bullet \;\;\; a \sim t^{2/3}$\\
$\bullet \;\;\; h \sim t^{-1}$\\
$\bullet \;\;\; f^2 \sim k t^{-4/3} + t^{-2}$.\\
Note that in this limit the scale factor ($a$) has a behaviour that resembles that of the standard FRWL cosmology for some particular values of $k$ and $\rho$. For each value of $\Lambda$, the closest matching cases are:
\begin{center}
$\begin{array}{c|c}
	\mbox{Here } C=0, \rho\neq 0 & \mbox{Standard FRWL}  \\ \hline
	\Lambda>0	& \Lambda>0, \rho=0, k=1 \\ \hline
	\Lambda<0 & \Lambda<0, \rho=0, k=-1 \\ \hline
	\Lambda=0 & \Lambda=0, \rho\neq 0, p=0, k=0
 \end{array}$
\end{center}
Even though the time evolution is qualitatively similar in each case, the right column can not be seen as a limit of the left column (because the latter does not depend on $k$ or $\rho$). In fact, the usual equations of cosmology can not be obtained as a limit of our model. The limit $f=h=0$ of (\ref{1}), (\ref{2}), (\ref{5}) implies $\dot{a}^2 = -k$ which only has solution for $k = 0$ (static universe) and $k = -1$ (linearly expanding universe) and $p = - \rho = \Lambda$. This limit does not correspond to any standard scenario. This is due to the fact that the Euler-scalar coupling is not a perturbative deformation of standard Einstein Hilbert theory. 

It can be noted that a solution of the form $H \sim tanh(\omega t)$ can also be obtained in very different theories. For example, in \cite{Aref'eva:2007yr} a solution like that is obtained in a General Relativity model with a phantom scalar field (without potential), for k = 0 and positive cosmological constant. A similar result is found in \cite{Biswas:2005qr}  in a nonlocal modified gravity model.

%%%%%%%%%%%%%%%%%%%%%%%%%%%%%%%
\section{Discussion and summary}   % 4 %
%%%%%%%%%%%%%%%%%%%%%%%%%%%%%%%
A minor variation of the form of coupling discussed here would be to change $\phi \rightarrow V(\phi)$. This only means that the function that couples to the Euler density is not a fundamental field, and offers some flexibility for the field equations because in the equations it replaces $\mathrm{d}\phi$ by $V'(\phi) \mathrm{d}\phi$. This results in different scenarios for $V'(\phi_0)=0$ and $V'(\phi_0) \neq 0$, where $\phi_0$ is a classical configuration. We have not discussed this case in detail here as it does not add a very significant conceptual difference to the previous discussion, although it may entail important cosmological consequences that deserve careful analysis. 

The specific nature of the field $\phi$ is not very important, its origin could be exotic or mundane, but the fact that it is present makes a huge difference. In an inhomogeneous and anisotropic universe, the scalar field could be interpreted as the effect of averaging over those local fluctuations \cite{Buchert:2006ya}. Another source for the type of coupling considered here is from the effective action derived by the anomaly of a classical scale-invariant field theory \cite{Gilkey:2002qd}.

One could try to refine the model by assuming a dynamical scalar field $\phi$. For instance, assuming matter to be described by the Klein-Gordon Lagrangian $L_m = \frac{1}{2} (\partial\phi)^2$, the density and pressure of the scalar field become $p = \rho = \frac{1}{2}\dot{\phi}^2$. The modified equations are completely integrable (see Apendix B for details). However, a numerical analysis beyond the scope of this work would be needed in order to obtain detailed information about the solutions.

Although our model might be regarded as not very realistic, it shows how a slight modification in the form of the action, having negligible effects in a small region (e.g., at the scale of the solar system at the present age of the universe), can give rise to dramatic changes in the cosmic evolution. Let us summarize the main features of this very simple scenario:\\

$\bullet$ For $\Lambda > 0$, the solution does not have an initial, point-like singularity. The universe starts from an infinitely large scale, contracting to a minimum size before starting an era of accelerated expansion.\\

$\bullet$ For $\Lambda < 0$, there is an initial point-like singularity as in the standard FRWL cosmology. At late times, the universe also experiences an accelerated expansion, reaching an infinitely large scale factor in a finite cosmic time, $t=\pi/\sqrt{-2\Lambda}$.\\

$\bullet$ For $\Lambda = 0$ the universe also starts with an initial singularity, but in this case there is no accelerated expansion, the universe, instead, contracts and collapses in a finite time, $t_c = \frac{\kappa^2\rho_0}{3 C}$, that is proportional to the present density of the universe .\\

$\bullet$ The time evolution near the big bang is highly sensitive to the sign of $\Lambda$. The solution for $\Lambda = 0$ cannot be obtained by taking the limit of $\Lambda \rightarrow 0$ from the $\Lambda \neq 0$ solutions, however its behaviour is very similar to the case $\Lambda<0$ near the big bang.\\

$\bullet$ The evolution of the scale factor is insensitive to the value of the curvature of the spatial section $k$. This means that in this scenario one could not infer the value of $k$ form the cosmological expansion, which in turn would make the flatness problem irrelevant since there is no need to explain the critical value of the energy density to match the observations.\\

$\bullet$ The universe can undergo an accelerated expansion for both signs of the cosmological constant, while in standard GR with $\Lambda< 0$, an accelerated expansion can only take place in the presence of exotic matter, or other exceptional conditions \cite{Prokopec}. \\

In this model, torsion was not included explicitly in the action, but indirectly introduced through a scalar field non-minimally coupled to gravity, a slight modification of the standard assumptions of General Relativity. Thanks to torsion (and the Euler-Scalar coupling), the universe undergoes an accelerated expansion without requiring either dark energy or negative pressure fluids. In fact, the nonvanishing torsion contributes to both ``matter density" (\ref{rho}) and ``pressure" (\ref{p}), which are not positive definite quantities. 

Apart from the Euler class, there exist other topological invariants like the Pontryagin and the Nieh-Yan\cite{Nieh-Yan,Mardones}  forms that could also be analogously coupled to a scalar field.  We are aware of several explorations in this direction,  \cite{Jackiw-Pi,Botta,Ertem,Cambiaso}. Membranes of co-dimension two or more can also couple to Chern-Simons forms \cite{Miskovic,Edelstein}, bringing in torsion as well.

As seen in Fig. (\ref{fsquare}) for $\Lambda>0$ and $t\rightarrow 0$,  as well as for $\Lambda\leq 0$ and $t\rightarrow \infty$,  $f^2$ becomes negative. This situation may seem catastrophic, but since $f$ corresponds to the completely antisymmetric part of the contorsion, it does not affect the geodesic equation of classical particles and does not couple to spin zero or gauge vectors fields. This component of the torsion only affects spin $\frac{1}{2}$ particles through the coupling in the Dirac Lagrangian \cite{Carroll-Field} 
\begin{equation}
\mathcal{L}_I = - \frac{i}{8} T_{\alpha \beta \gamma} \bar{\psi} \Gamma^{\alpha \beta \gamma} \psi = \frac{3}{2} f \bar{\psi} \Gamma^0 \Gamma^5 \psi \; .
\end{equation}
 This interaction term couples torsion to the chiral current and is added to the torsion-free Dirac Lagrangian $\bar{\mathcal{L}}_D = \frac{i}{2}(\bar{\psi} \Gamma^\mu \bar{\mathrm{D}}_\mu \psi - \bar{\mathrm{D}}_\mu \bar{\psi} \Gamma^\mu \psi) - i m \bar{\psi}\psi$. If $f$ becomes imaginary, the corresponding loss of unitarity and violation of the current conservation can be interpreted as giving rise to particle creation.
 
Among all the possible solutions discussed above, the case $\Lambda <0$ and $k=0$ seems closest to the present cosmological evidence: a big-bang singularity with an accelerated expansion and a negligible presence of torsion effects at late times. 

The current limited understanding of the evolution of the universe at large scales urges a revision of the standard assumptions of cosmology. The case presented here is just one example of how dramatically different are the resulting scenarios if certain logically unnecessary assumptions we take for granted --like the absence of torsion or the metric formulation of GR--, are removed. Another example where this kind of assumptions are removed is presented in \cite{Poplawski}.

%%%%%%%%%%%%%%%%
\section*{Appendix A} 
%%%%%%%%%%%%%%%%
With the convenient choice of the vierbein $e^0=\mathrm{d}t$ and $e^1 = \frac{a(t)}{\sqrt{1-kr^2}}\mathrm{d}r$, $e^2 = a(t) r \mathrm{d}\theta $, $e^3 = a(t) r \sin \theta \mathrm{d}\varphi$, the torsion 2-form becomes\footnote{Here $I,J,K=\{1,2,3\}$ are Lorentz indices.}
\begin{eqnarray}
\label{T0} T^0 &=& 0\\
\label{Ti} T^I &=& h(t) e^I e^0 + f(t) {\epsilon^I}_{JK} e^J e^K \, ,
\end{eqnarray}
From the equation $T^a = {\kappa^a}_b e^b$ we identify the contorsion one-form to be $\kappa^{0I} = h(t) e^I$ and $\kappa^{IJ} = -f(t) \epsilon^{IJ}{}_K e^K$. On the other hand, the nonvanishing components of the torsionless part of the spin connection are $\bar\omega^{0I} = H e^I$, $\bar\omega^{12} = - (a r)^{-1} e^2\sqrt{1-kr^2}$, $\bar\omega^{13} =  - (a r)^{-1} e^3\sqrt{1-kr^2}$ and $\bar\omega^{23} = - (a r)^{-1} e^3\cot \theta $; where $H = \dot{a}/a$  is the Hubble function, and therefore the spin connection reads
\begin{eqnarray}
\label{w0i} \omega^{0I} &=& (H+h) e^I\\
\label{w12} \omega^{12} &=& -\frac{\sqrt{1-kr^2}}{a r} e^2 - f e^3 \\
\label{w13} \omega^{13} &=& -\frac{\sqrt{1-kr^2}}{a r} e^3 + f e^2 \\
\label{w23} \omega^{23} &=& -\frac{\cot \theta}{a r} e^3 - f e^1
\end{eqnarray}
The Lorentz curvature $R^{ab} = \mathrm{d} \omega^{ab} + {\omega^a}_c \omega^{cb}$ takes the form
\begin{eqnarray}
\label{R0i} R^{0I} &=& [(\dot{H} + \dot{h}) + H (H + h)] e^0 e^I + f (H + h) {\epsilon^I}_{JK} e^J e^K \\
\label{Rij} R^{IJ} &=& \Big[(H + h)^2 + \frac{k}{a^2} - f^2\Big] e^I e^J + (\dot{f} + H f) {\epsilon^{IJ}}_K e^K e^0
\end{eqnarray}
The stress 3-Form is related to the stress tensor by $\tau_d = \frac{1}{6} \epsilon_{labc} {\tau^l}_{d} e^a e^b e^c$, and considering a perfect fluid ${\tau^l}_{d} = diag(-\rho,p,p,p)$ we obtain
\begin{eqnarray}
\label{stress0} \tau_0 &=& -\frac{1}{6} \rho\ \epsilon_{IJK} e^I e^J e^k \\
\label{stressI} \tau_I &=& -\frac{1}{2} p\ \epsilon_{IJK} e^0 e^J e^k
\end{eqnarray}
Finally, replacing (\ref{T0}, \ref{Ti}, \ref{R0i}, \ref{Rij}) into the field equations (\ref{Einstein})-(\ref{Euler}), one obtains Eqs.(\ref{1})-(\ref{5}).\\

%%%%%%%%%%%%%%%%
\section*{Appendix B}
%%%%%%%%%%%%%%%%

One way to refine the model is to assume a dynamical scalar field $\phi$. In that case, the matter Lagrangian is $L_m(\phi) = \frac{1}{2}\mathrm{d}\phi * \mathrm{d}\phi$. The stress tensor is $\tau_{ab} = \partial_a \phi \partial_b \phi - \frac{1}{2} \eta_{ab} \partial_c \phi \partial^c \phi$ (where we defined $\partial_a = E^\alpha_a \partial_\alpha$), and since $\phi=\phi(t)$, the density and pressure of the scalar field are given by $p = \rho = \frac{1}{2}\dot{\phi}^2$. In this case, (\ref{Euler}) is also modified as $\epsilon_{abcd} R^{ab} R^{cd} = 4 \kappa^2 d*d\phi$. Assuming again that $f\neq 0$, (\ref{6}) is replaced by
\begin{equation}
\label{7} \frac{U^3}{3} + U\Big(\frac{k}{a^2} - f^2\Big) - \frac{\kappa^2}{12 U} = \frac{C}{a^3}
\end{equation}
Substituting (\ref{1}) in (\ref{2}, \ref{3}) again reduces the problem to quadratures that can be integrated for $U(t)$. If $|\Lambda| \neq \kappa$, then 
\begin{equation*}
 t-t_0 = \sqrt{\frac{-\Lambda-\sqrt{\Lambda^2-\kappa^2}}{\Lambda^2- \kappa^2}}\arctan\bigg[\frac{2 U}{\sqrt{-\Lambda-\sqrt{\Lambda^2-\kappa^2}}}\bigg] - \sqrt{\frac{-\Lambda+\sqrt{\Lambda^2-\kappa^2}}{\Lambda^2-\kappa^2}}\arctan\bigg[\frac{2 U}{\sqrt{-\Lambda+\sqrt{\Lambda^2-\kappa^2}}}\bigg].
\end{equation*}
The cases $\Lambda =\pm \kappa$ must be handled separately, giving
\begin{eqnarray}
\nonumber t - t_0 = \frac{2 U}{4 U^2 - \kappa} + \frac{\tanh^{-1}\Big[\frac{2 U}{\sqrt{\kappa}}\Big]}{\sqrt{\kappa}}  & \&  & t - t_0 =  \frac{2 U}{4 U^2 + \kappa} - \frac{\arctan\Big[\frac{2 U}{\sqrt{\kappa}}\Big]}{\sqrt{\kappa}} \,\,,
\end{eqnarray}
respectively. For $\Lambda = 0$ the solution is
\begin{equation*}
\sqrt{2 \kappa} (t-t_0) = \arctan \Big[1-\sqrt{\frac{8}{\kappa}}U\Big] - \arctan \Big[1+\sqrt{\frac{8}{\kappa}}U\Big] + \frac{1}{2} \log \bigg[\frac{(\sqrt{8} U + \sqrt{\kappa})^2 + \kappa}{(\sqrt{8} U - \sqrt{\kappa})^2 + \kappa} \bigg] \, . 
\end{equation*}
Finally, combining this with Eqs. (\ref{1}), (\ref{3}), (\ref{7}) and $\dot{\phi}=\frac{1}{2U}$, the system can be completely integrated.\\

\textbf{\Large Acknowledgments}\newline
Enlightening discussions with J.~S.~Alcaniz, M.~Cambiaso, S.~del~Campo, M.~Hassaine, R.~Herrera, G.~Leon, O.~Mi\v{s}kovi\'c, J.~Pacheco, M.~Tsoukalas, P.~Vargas-Moniz, A.~Zelnikov and T.~Zlosnik are warmly acknowledged. This work was supported by Fondecyt grants \# 1110102, 1100328, 1100755, 1100328, and by Conicyt grant \textit{Southern Theoretical Physics Laboratory, ACT-91}. A.T. acknowledges financial support by program MECESUP 0605. The Centro de Estudios Cient\'{\i}ficos (CECS) is funded by the Chilean Government through the Centers of Excellence Base Financing Program of Conicyt.

\end{document}